# Conservation of Total Escape from Hydrodynamic Planetary Atmospheres

Feng Tian[1,2]

1. National Astronomical Observatories, Chinese Academy of Sciences

2. Center for Earth System Sciences, Tsinghua University

**Abstract: Atmosphere escape is one key process controlling the evolution of planets. However, estimating the escape rate in any detail is difficult because there are many physical processes contributing to the total escape rate. Here we show that as a result of energy conservation the total escape rate from hydrodynamic planetary atmospheres where the outflow remains subsonic is nearly constant under the same stellar XUV photon flux when increasing the escape efficiency from the exobase level, consistent with the energy limited escape approximation. Thus the estimate of atmospheric escape in a planet's evolution history can be greatly simplified.**

## 1. Introduction

Recently it is proposed that a high hydrogen content (at least a few percent) in early Earth's atmosphere could be important to keep early Earth warm, contributing to the solution of the faint young Sun problem (Wordsworth and Pierrehumbert 2013). A hydrogen rich early Earth atmosphere has been proposed based on hydrodynamic calculations of hydrogen escape and such an atmosphere could have been important for prebiotic photochemistry (Tian et al. 2005a). The numerical scheme in Tian et al. (2005a) contains large numerical diffusion especially near the lower boundary where the density gradient is the largest (Tian et al. 2005b). However, the calculated upper atmosphere structures are consistent with transit observations of hot Jupiter HD209458b (Vidar-Madjar et al. 2003) and the calculated escape rates are consistent with follow-up independent works (Yelle 2006, Garcia-Munoz 2007, Penz et al. 2008, Koskinen et al. 2012). On the other hand, it is suggested that nonthermal escape processes could have increased total hydrogen escape (Catling 2006) so that the hydrogen content in early Earth atmosphere would have been in the order of 0.1% instead of a few percent or greater. But no calculation has been carried out to estimate the nonthermal hydrogen escape rate from early Earth's atmosphere.

Lammer et al. (2007) showed that planets in the habitable zones of M dwarfs should experience frequent exposure to stellar corona mass ejection events and as a result Earth-like planets in such environments could have lost hundreds of bars of $CO_2$ in the timescale of 1 Gyrs through stellar wind interactions. But the energy consumed in such massive atmospheric loss is not considered. Tian (2009) showed that $CO_2$-dominant atmospheres of super Earths with masses greater than 6 Earth masses should survive the long active phase of M dwarfs. But nonthermal escape processes were not included.

Observed close-in exoplanets with masses in the range between Earth and Uranus/Neptune, such as Corot-7B (Leger et al. 2009), GJ1214b (Charbonneau 2009), 55 Cnc e (Winn et al. 2011), and several Kepler-11 planets (Lissauer et al. 2011), have inferred densities much different from each other. Considering the small orbital distances between these planets and their parent stars, these planetary atmospheres must be highly expanded and atmospheric escape must be an important physical process controlling the evolution histories and the nature of these objects. A better understanding of the relationship between different atmospheric escape processes and how upper atmosphere structure influences atmospheric escape is urgently needed.

One classical theory on atmospheric escape is the diffusion-limited escape (Hunten 1973), which provides an upper limit for the total escape rate of minor species. In the case of hydrogen, its escape rate should be proportional to the total mixing ratio of hydrogen-bearing species at the homopause level. The diffusion-limited escape theory is the result of the kinetics at the homopause level and does not consider the energy aspect of atmospheric escape.

When a planetary atmosphere is exposed to intense stellar XUV photon flux, which occurs on terrestrial planets during their early evolution histories, close-in exoplanets, and small dwarf planets such as present Pluto, the upper atmosphere is heated and temperature rises and the atmosphere expands. For this scenario to occur, thermal conduction through the lower boundary must be less than the net heating. When the atmosphere expands to large distance, the gravity of the planet at the exobase, the top of the atmosphere, becomes weak enough and major atmospheric species escape more efficiently through either thermal or nonthermal processes. When the escape of major atmospheric species is efficient, the upper atmosphere flows outward and the adiabatic cooling associated with the expansion of the rapidly escaping atmosphere becomes a dominant part of the energy budget of planetary atmospheres -- the hydrodynamic regime or a hydrodynamic planetary atmosphere (Tian et al. 2008a, b). Because the diffusion-limited theory does not consider energy required to support rapid escape, it cannot provide us a good estimate on escape rate of major atmospheric species

Note that there is a difference between the above-mentioned hydrodynamic planetary atmosphere and the traditional hydrodynamic escape, or blowoff, in that the hydrodynamic regime is reached when the outflow is important in the energy budget of the upper atmosphere, while the blowoff occurs when the heating of the upper atmosphere is so strong that the kinetic energy of the upper atmosphere overcomes the gravity of the planet. Thus a planetary atmosphere in the hydrodynamic regime does not necessarily blow off. In such an atmosphere the gravitational potential energy is more than the heat content or kinetic energy of the atmosphere and the atmospheric escape is Jeans-like (evaporation) no matter whether the actual escape process is thermal or nonthermal. Thus a planetary atmosphere could be experiencing Jeans-like escape and in the hydrodynamic regime simultaneously (Tian et al. 2008a). On the other hand, blowoff can be considered an extreme case of planetary atmospheres in the hydrodynamic regime and energy consumption in the outflow is the ultimate factor controlling the mass loss rate.

Linking the hydrogen content of early Earth's atmosphere with the nature of close-in super Earths, the key question this paper intends to address is: can the energy requirement in a hydrodynamic planetary atmosphere limit atmospheric escape?

## 2. Hydrodynamic Planetary Upper Atmospheres and the Conservation of Total Escape Rate

Here a 1-D upper planetary atmosphere model, validated against the upper atmosphere of the present Earth, is used to study the problem. The model details can be found in Tian et al. (2008a, b). A key feature of the model is that it can automatically adjust its upper boundary so that the exobase, defined as where the scale height is comparable to the mean free path, can be found and the adjusted Jeans escape rates of all species can be calculated. When increasing the level of solar XUV radiation, both the upper atmosphere temperature and the exobase altitude increase. At 5 times present solar mean XUV level (XUVx5), the exobase altitude can reach more than $10^4$ km and the upper atmosphere temperature can be near 9000 K (Tian et al. 2008b).

To include other escape processes at the exobase level in addition to Jeans escape, the Jeans escape effusion velocity at the exobase is multiplied by 3, 10, and 20 times respectively. The calculated upper atmosphere temperature profiles are shown in Fig. 1. The peak temperature in the upper atmosphere cools with increasing escape efficiency from 9000 K in the Jeans escape only case to 8000, 7500, and 7000 K in the 3x, 10x, and 20x more efficient atmosphere escape cases. Correspondingly the exobase altitude decreases with increased escape efficiency because of decreased scaleheight. Note that although the scaleheight is inversely proportional to the temperature, the exobase altitude is not.

The shrinking of the upper atmosphere with increasing escape efficiency at the exobase level has an interesting consequence on the total atmospheric escape rate, shown as a solid curve in Fig. 2. In comparison the dashed line in Fig. 2 shows a linear increase of total escape with enhanced escape efficiency if the upper atmosphere structure is not influenced by atmospheric escape. When considering the energy required to support a strong outflow, which is a consequence of rapid escape of major atmosphere species, the total escape rate of such species remains almost a constant (a conservation of total escape rate) when increasing escape efficiency from the exobase level. The conservation of total escape rate from a hydrodynamic planetary atmosphere is a demonstration of the law of the conservation of energy -- changing the escape efficiency at the exobase level does not change the total amount of energy heating the upper atmosphere.

## 3. Discussion

The heating and cooling terms in the hydrodynamic planetary atmosphere are shown in Fig. 3. The solid curves correspond to the black curve in Fig. 1 (Jeans escape only). The dashed curves correspond to the red curve in Fig. 1 (escape efficiency 10 times that of Jeans escape).

The blue curves represent the adiabatic cooling. The red curves represent the net radiative heating, which includes absorption of stellar XUV photons, the ionization, excitation, and dissociation of atmospheric species, and transport of energetic electrons and the deposition of their energy, heating from chemical reactions, and radiative cooling (for details see Tian et al. 2008a, b). The magenta curves represent the thermal conduction.

Fig. 3 shows that the adiabatic cooling associated with outflow is a dominant cooling term in the upper thermosphere and its importance increases with altitude, reflecting the increasing velocity of the outflow. Although thermal conduction cooling is dominant near where the net radiative heating peaks, because the net radiative heating is inadequate to match the adiabatic cooling, thermal conduction becomes an important heating term in the upper thermosphere, contributing to support the outflow. When increasing the escape efficiency at the exobase level, the outflow is enhanced in upper thermosphere, which causes the dashed blue curve to increase more rapidly with altitude than the solid blue curve does. The peak of the net radiative heating moves lower in altitude because the atmosphere shrinks, which allows more stellar XUV photons to penetrate to deeper altitudes. Fig. 3 and Fig. 1 show how a planetary upper atmosphere in the hydrodynamic regime adjusts its structure and energy distribution when different escape processes occur at the exobase level. As a result of the law of the conservation of energy, the total escape rates from the two atmospheres with quite different structures are almost identical.

We further tested the 1-D upper atmosphere model with greater XUV levels and in all cases the enhancements of atmospheric escape efficiency lead to shrink of the upper atmosphere and the conservation of total escape rate. For an upper atmosphere under weak XUV heating, the escape is insignificant and the increased escape efficiency does not lead to a strong outflow or cooling of the upper atmosphere. Thus the total escape increases linearly with enhanced escape efficiency. Analysis shows that the difference between the two cases is whether the cooling caused by the gas outflow is important in the energy budget of the upper atmosphere, which is the division between a hydrostatic and a hydrodynamic planetary upper atmosphere.

Johnson et al. (2013) pointed out that if the deposition of energy in the upmost layer of the thermosphere (where the ratio between the mean free path and the scaleheight is >~0.1) is inefficient, nonthermal escape processes can be ignored. This is equivalent of saying that the energy deposited in the collision-dominant part of the atmosphere contributes to the heating of the atmosphere and not to nonthermal escape processes. Our model atmospheres include this heating process. From the perspective of the exobase, the escape is Jeans-like in that the bulk outflow velocity in our model remains subsonic, consistent with the findings in Johnson et al. (2013). From the perspective of the energy budget of the upper atmosphere, the escape is hydrodynamic because the outflow cooling dominates the energy budget when the atmosphere is under strong XUV radiation (Tian et al. 2008a). Thus escape can be Jeans-like and hydrodynamic simultaneously, depending on from which point of view the issue is observed.

Garcia-Munoz (2007) showed that the model calculated escape rates from HD209458b are insensitive to the upper boundary conditions but the upper atmospheric structures are. Koskinen et al. (2012) compared the effects of different boundary conditions on the escape from hot Jupiters and found that models with similar escape rates could produce different upper atmospheric structures depending on the boundary conditions applied. These results are in good agreement with ours and thus the conservation of total escape rate theory proposed in this paper is also supported by models with transonic hydrogen outflow from hot Jupiters. In agreement with Koskinen et al. (2012), we emphasize that details of the escape processes functioning at the exobase level are important for understanding the upper atmosphere structures and thus are important to compare with observations.

A recent hybrid fluid/kinetic model for the upper atmosphere of Pluto (Erwin et al. 2013) shows that Pluto's exobase altitude and temperature decreases as a result of increasing escape efficiency at the exobase level and as a result the total escape rate from Pluto's $N_2$-dominant atmosphere remains near constant.  This shows that the theory of the conservation of total escape rate applies to hydrodynamic planetary atmosphere with different composition.  Erwin et al. (2013) pointed out that in order to predict the upper atmosphere structure, a hybrid fluid/kinetic model is needed because the enhancement of escape efficiency at the exobase level does influence the upper atmosphere structure. We emphasize that if the theory of the conservation of total escape is correct, a fluid model for the planetary upper atmosphere would be adequate to understand the atmospheric escape history of a planet.

The conservation of total escape rate from planetary atmospheres in the hydrodynamic regime is demonstrated in 1-D models for the Earth's current atmosphere composition under intense XUV heating, for the current $N_2$-dominant atmosphere of Pluto (Erwin et al. 2013), and for transonic outflow from hot Jupiters (Garcia-Munoz 2007, Koskinen et al.

2012). Numerical simulations for atmospheres with different composition around planets with different masses in 3-D models will be needed in future studies to prove or disprove it. However we speculate that the theory should apply to planetary atmospheres with different composition because the law of the conservation of energy is universal and the escape of hydrogen from such atmospheres under intense stellar/solar XUV heating is always energy-limited.

If the theory is confirmed, one implication is that early Earth's atmospheric hydrogen content should be close to those in Tian et al. (2005) suggested, provided that those calculations are correct, and thus hydrogen could have helped early Earth to stay warm (Wordsworth and Pierrehumbert 2013). And the calculations of atmosphere escape during the evolution histories of different planets could be significantly simplified. The theory also implies that super Earths in close-in orbits can have a better chance to keep their atmospheres and oceans, which might help to explain the existence of low density rocky exoplanets such as 55 Cnc e and GJ1214b, and planets in the habitable zones of M dwarfs should be able to keep their $CO_2$-dominant atmospheres, supporting the conclusion of Tian (2009), which could have consequences in the evaluation of planetary habitability.

## 4. Conclusions

As a result of the law of the conservation of energy, the total escape rate from planetary atmospheres in the hydrodynamic regime is nearly constant under the same stellar XUV photon flux when increasing the escape efficiency from the exobase level. Thus an energy-limited escape approximation can be applied to such atmospheres, provided that the upper atmosphere structures are calculated accurately. The estimate of atmospheric escape in a planet's evolution history can be greatly simplified.

**Acknowledgement**: The author thanks R.E. Johnson and the other anonymous reviewer for their helpful comments and suggestions.

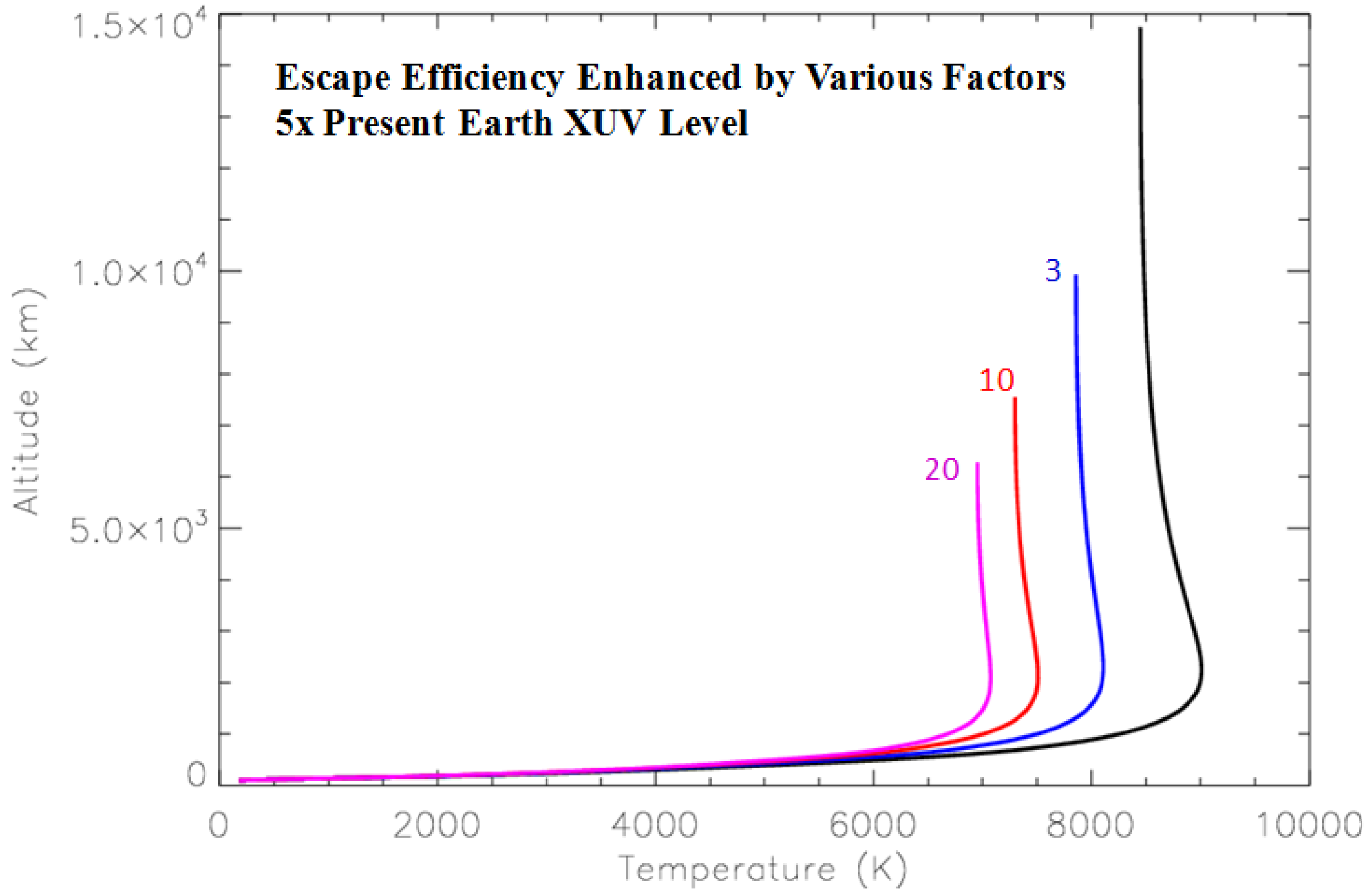

Fig. 1 Upper atmosphere structures of the Earth under 5 times present XUV radiation level with different escape effusion velocities at the exobase level, which are where the curves end.

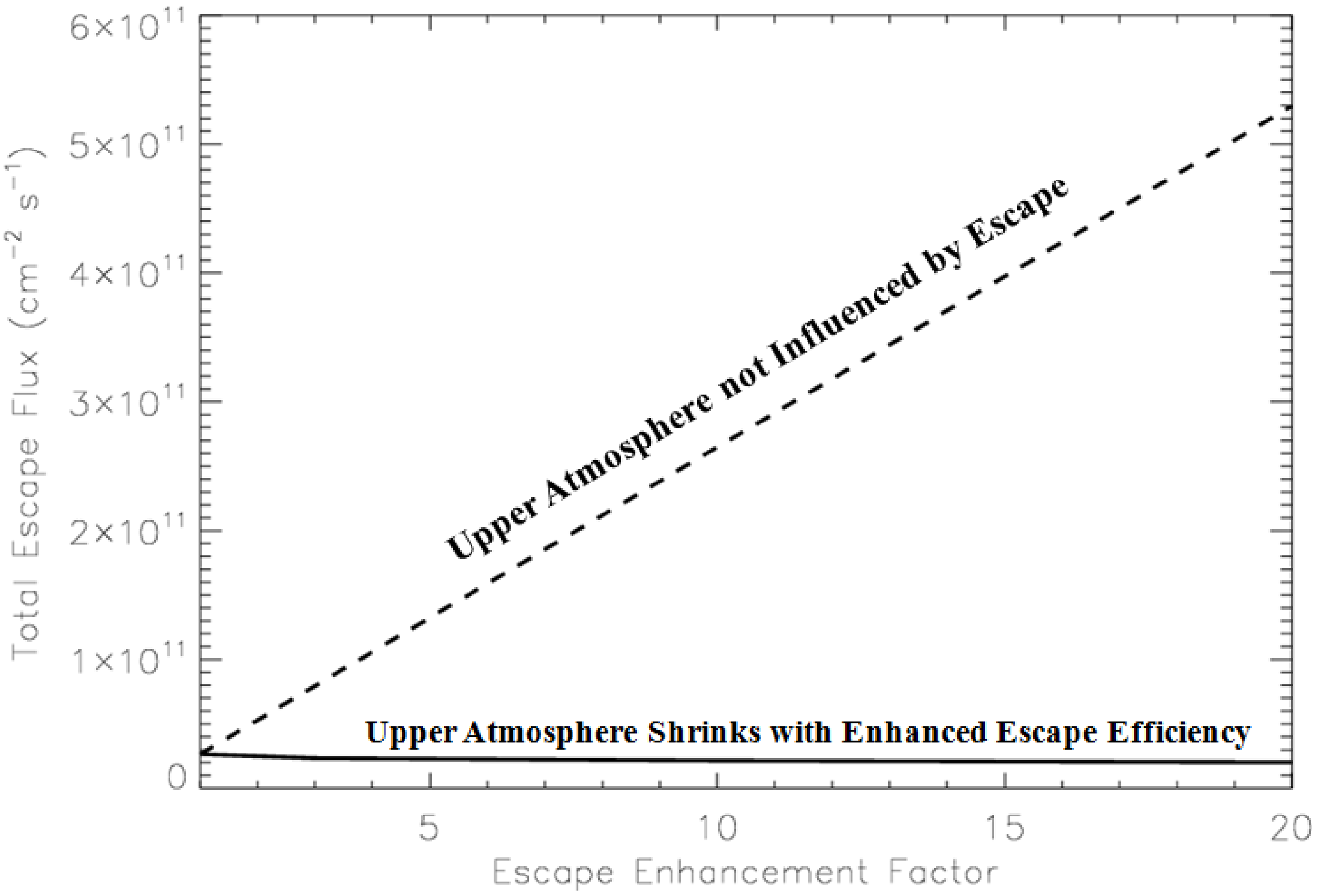

Fig. 2 Total escape rate of major atmosphere species as a function of escape efficiency from the exobase level. The atmospheres used in these simulations have composition the same as that of present Earth but are under 5 times present Earth's XUV radiation level. If the upper atmosphere structure is not influenced by escape of major atmospheric species and the subsequent outflow, the total escape rate would have increased linearly with enhanced escape efficiency at the exobase level as shown by the dashed line. However, when considering the energy consumption of outflow in the upper atmosphere, the upper atmosphere cools and shrinks (shown in Fig. 1) and the total escape rate remains conserved with enhanced escape efficiency at the exobase level.

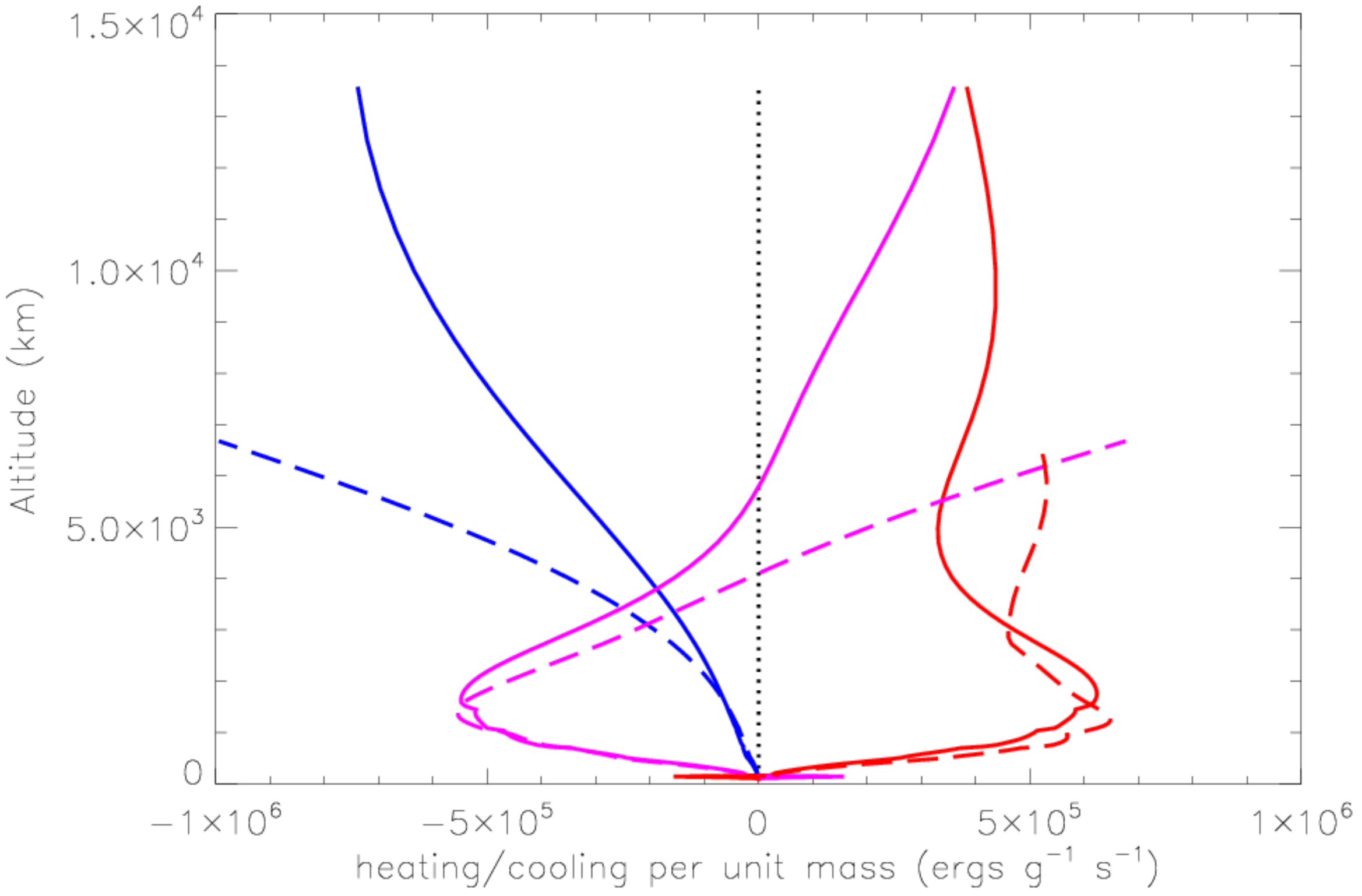

Fig. 3 Heating and cooling terms in a hydrodynamic planetary atmosphere. The solid curves correspond to the black curve in Fig. 1. The dashed curves correspond to the red curve in Fig. 1. The blue curves are the adiabatic cooling. The red curves are the net radiative heating. The magenta curves are the thermal conduction.